 \documentclass[final,3p,times,twocolumn]{elsarticle}

\usepackage{amssymb}
\usepackage{url}
\usepackage{lineno}
\usepackage{subfig}
\usepackage{wrapfig}
\usepackage{booktabs}
\usepackage{hyperref}
\usepackage[utf8]{inputenc}

\journal{Nuclear Inst. and Methods in Physics Research, A}

\begin{document}

\begin{frontmatter}



\title{\normalsize \textbf{In-beam performance of a Resistive Plate Chamber operated with eco-friendly gas mixtures}}




\author[1]{\small L. Quaglia}
\author[6,3]{\small on behalf of the ALICE and RPC ECOgas@GIF++ collaborations. \textbf{The RPC ECOgas@GIF++ collaboration: }M. Abbrescia}
\author[21]{\small G. Aielli}
\author[3,8]{\small R. Aly}
\author[17]{\small M. C. Arena}
\author[11]{\small M. Barroso}
\author[4]{\small L. Benussi}
\author[4]{\small S. Bianco}
\author[7]{\small D. Boscherini}
\author[16]{\small F. Bordon}
\author[7]{\small A. Bruni}
\author[18]{\small S. Buontempo}
\author[16]{\small M. Busato}
\author[21]{\small P. Camarri}
\author[13]{\small R. Cardarelli}
\author[3]{\small L. Congedo}
\author[11]{\small D. De Jesus Damiao}
\author[6,3]{\small M. De Serio}
\author[21]{\small A. Di Ciaccio}
\author[21]{\small L. Di Stante}
\author[14]{\small P. Dupieux}
\author[19]{\small J. Eysermans}
\author[12,1]{\small A. Ferretti}
\author[6,3]{\small G. Galati}
\author[12,1]{\small M. Gagliardi}
\author[16]{\small R. Guida}
\author[2,3]{\small G. Iaselli}
\author[14]{\small B. Joly}
\author[24]{\small S.A. Juks}
\author[23]{\small K.S. Lee}
\author[13]{\small B. Liberti} 
\author[22]{\small D. Lucero Ramirez}
\author[16]{\small B. Mandelli}
\author[14]{\small S.P. Manen}
\author[7]{\small L. Massa}
\author[3]{\small A. Pastore}
\author[13]{\small E. Pastori}
\author[4]{\small D. Piccolo}
\author[13]{\small L. Pizzimento}
\author[7]{\small A. Polini}
\author[13]{\small G. Proto}
\author[2,3]{\small G. Pugliese}
\author[2,3]{\small D. Ramos}
\author[16]{\small G. Rigoletti}
\author[13]{\small A. Rocchi}
\author[7]{\small M. Romano}
\author[10]{\small A. Samalan}
\author[9]{\small P. Salvini}
\author[21]{\small R. Santonico}
\author[5]{\small G. Saviano}
\author[13]{\small M. Sessa}
\author[6,3]{\small S. Simone}
\author[12,1]{\small L. Terlizzi}
\author[10,20]{\small M. Tytgat}
\author[12,1]{\small E. Vercellin}
\author[15]{\small M. Verzeroli}
\author[22]{\small N. Zaganidis}

\affiliation[1]{organization={INFN Sezione di Torino},
             addressline={Via P. Giuria 1},
             city={Torino},
             postcode={10125},
             state={Italy},
             country={}}
             
\affiliation[2]{organization={Politecnico di Bari, Dipartimento Interateneo di Fisica},
             addressline={via Amendola 173},
             city={Bari},
             postcode={70125},
             state={Italy},
             country={}}

\affiliation[3]{organization={INFN Sezione di Bari},
             addressline={Via E. Orabona 4},
             city={Bari},
             postcode={70125},
             state={Italy},
             country={}}

\affiliation[4]{organization={INFN - Laboratori Nazionali di Frascati},
             addressline={Via Enrico Fermi 54},
             city={Frascati (Roma)},
             postcode={00044},
             state={Italy},
             country={}}

\affiliation[5]{organization={Sapienza Università di Roma, Dipartimento di Ingegneria Chimica Materiali Ambiente},
             addressline={Piazzale Aldo Moro 5},
             city={Roma},
             postcode={00185},
             state={Italy},
             country={}}

\affiliation[6]{organization={Università degli studi di Bari, Dipartimento Interateneo di Fisica},
             addressline={Via Amendola 173},
             city={Bari},
             postcode={70125},
             state={Italy},
             country={}}

\affiliation[7]{organization={INFN Sezione di Bologna},
             addressline={Via C. Berti Pichat 4/2},
             city={Bologna},
             postcode={40127},
             state={Italy},
             country={}}

\affiliation[8]{organization={Helwan University},
             addressline={},
             city={Helwan, Cairo Governorate},
             postcode={4037120},
             state={Egypt},
             country={}}

\affiliation[9]{organization={INFN Sezione di Pavia},
             addressline={Via A. Bassi 6},
             city={Pavia},
             postcode={27100},
             state={Italy},
             country={}}

\affiliation[10]{organization={Ghent University, Dept. of Physics and Astronomy},
             addressline={Proeftuinstraat 86},
             city={Ghent},
             postcode={B-9000},
             state={Belgium},
             country={}} 
             
\affiliation[11]{organization={Universidade do Estado do Rio de Janeiro},
             addressline={R. São Francisco Xavier, 524},
             city={Maracanã, Rio de Janeiro - RJ},
             postcode={20550-013},
             state={Brazil},
             country={}} 

\affiliation[12]{organization={Università degli studi di Torino, Dipartimento di Fisica},
             addressline={Via P. Giuria 1},
             city={Torino},
             postcode={10125},
             state={Italy},
             country={}} 

\affiliation[13]{organization={INFN Sezione di Roma Tor Vergata},
             addressline={Via della Ricerca Scientifica 1},
             city={Roma},
             postcode={00133},
             state={Italy},
             country={}}

\affiliation[14]{organization={Clermont Université, Université Blaise Pascal, CNRS/IN2P3, Laboratoire de Physique Corpusculaire},
             addressline={BP 10448},
             city={Clermont-Ferrand},
             postcode={F-63000},
             state={France},
             country={}}

\affiliation[15]{organization={Universitè Claude Bernard Lyon I},
             addressline={43 Bd du 11 Novembre 1918},
             city={Villeurbanne},
             postcode={69100},
             state={France},
             country={}}

 \affiliation[16]{organization={CERN},
             addressline={Espl. des Particules 1},
             city={Meyrin},
             postcode={1211},
             state={Switzerland},
             country={}}   
             
\affiliation[17]{organization={Università degli studi di Pavia},
             addressline={Corso Strada Nuova 65},
             city={Pavia},
             postcode={27100},
             state={Italy},
             country={}} 

\affiliation[18]{organization={INFN Sezione di Napoli},
             addressline={Complesso universitario di Monte S. Angelo ed. 6 Via Cintia},
             city={Napoli},
             postcode={80126},
             state={Italy},
             country={}}

\affiliation[19]{organization={Massachusetts Institute of Technology},
             addressline={77 Massachusetts Ave},
             city={Cambridge, MA},
             postcode={02139},
             state={USA},
             country={}}

\affiliation[20]{organization={Vrije Universiteit Brussel (VUB-ELEM), Dept. of Physics},
             addressline={Pleinlaan 2},
             city={Brussels},
             postcode={1050},
             state={Belgium},
             country={}}

\affiliation[21]{organization={Università degli studi di Roma Tor Vergata, Dipartimento di Fisica},
             addressline={Via della Ricerca Scientifica 1},
             city={Roma},
             postcode={00133},
             state={Italy},
             country={}}

\affiliation[22]{organization={Universidad Iberoamericana, Dept. de Fisica y Matematicas},
             addressline={},
             city={Mexico City},
             postcode={01210},
             state={Mexico},
             country={}}

\affiliation[23]{organization={Korea University},
             addressline={145 Anam-ro},
             city={Seongbuk-gu, Seoul},
             postcode={},
             state={Korea},
             country={}}

\affiliation[24]{organization={Université Paris-Saclay},
             addressline={3 rue Joliot Curie, Bâtiment Breguet},
             city={Gif-sur-Yvette},
             postcode={91190},
             state={France},
             country={}}

\begin{abstract}

\small ALICE (A Large Ion Collider Experiment) studies the Quark-Gluon Plasma (QGP): a deconfined state of nuclear matter obtained in ultra-relativistic heavy-ion collisions. One of the key probes for QGP characterization is the study of quarkonia and open heavy flavour production, of which ALICE exploits the muonic decay. In particular, a set of Resistive Plate Chambers (RPCs), placed in the forward rapidity region of the ALICE detector, is used for muon identification purposes.

\small The correct operation of these detectors is ensured by the choice of the proper gas mixture. Currently they are operated with a mixture of C$_{2}$H$_{2}$F$_{4}$, i-C$_{4}$H$_{10}$ and SF$_{6}$ but, starting from 2017, new EU regulations have enforced a progressive phase-out of C$_{2}$H$_{2}$F$_{4}$ because of its large Global Warming Potential (GWP), which is making it difficult and costly to purchase. Moreover, CERN asked LHC experiments to reduce greenhouse gases emissions, to which RPC operation contributes significantly.

\small A possible candidate for C$_{2}$H$_{2}$F$_{4}$ replacement is the C$_{3}$H$_{2}$F$_{4}$ (diluted with other gases, such as CO$_{2}$), which has been extensively tested using cosmic muons. Promising gas mixtures have been devised; the next crucial steps are the detailed in-beam characterization of such mixtures as well as the study of their performance under increasing irradiation levels.

\small This contribution will describe the methodology and results of beam tests carried out at the CERN Gamma Irradiation Facility (equipped with a high activity $^{137}$Cs source and muon beam) with an ALICE-like RPC prototype, operated with several mixtures with varying proportions of CO$_{2}$, C$_{3}$H$_{2}$F$_{4}$, i-C$_{4}$H$_{10}$ and SF$_{6}$ . Absorbed currents, efficiencies, prompt charges, cluster sizes, time resolutions and rate capabilities will be presented, both from digitized (for detailed shape and charge analysis) and discriminated (using the same front-end electronics as employed in ALICE) signals.

\end{abstract}


\end{frontmatter}


\section{Introduction}
\label{sec:intro}

\vspace{-10pt}

A Large Ion Collider Experiment (ALICE) \cite{ALICEtdr} is one of the four main LHC experiments. Its main goal is the study of the Quark Gluon plasma (QGP) in heavy-ion collisions. One of the exploited signatures for QGP studies in ALICE is the suppression of quarkonia (bound states of heavy quarks and corresponding anti-quark) production \cite{Jpsi}. ALICE studies their dimuon decays using its forward muon spectrometer \cite{muonTDR} and the Resistive Plate Chambers (RPCs), main subject of this contribution, are employed to identify the muons among all the particles crossing the spectrometer since they are located downstream of a hadron filter which absorbs any hadron traversing the muon spectrometer (hence the name Muon IDentifier, MID)  

RPCs are gaseous detectors with planar geometry and resistive electrodes. In the case of the ALICE MID, they have a 2~mm single gas gap and 2~mm thick bakelite electrodes ($\rho\approx$10$^{9}$-10$^{10}$~$\Omega\cdot$cm); they are readout by two perpendicular copper strips planes and they employ the FEERIC front-end electronics \cite{FEERIC}. The RPCs are operated in maxi-avalanche mode, with a mixture of 89.7\%~C$_{2}$H$_{2}$F$_{4}$, 10\%~i-C$_{4}$H$_{10}$ and 0.3\%~SF$_{6}$ and, although this mixture satisfies all the performance requirements, it contains a high fraction of fluorinated greenhouse gases (F-gases/GHGs) (C$_{2}$H$_{2}$F$_{4}$ and SF$_{6}$).

Starting from 2017, new EU regulations \cite{euReg} have imposed a progressive phase-down in the production and usage of these gases, leading to an increase of cost and reduction in availability. For this reason, CERN has adopted a policy of F-gases reduction. Since RPCs represent a significant fraction of the total GHG-gases emission of the LHC experiments, it is of the utmost importance to search for more eco-friendly RPC gas mixtures. The first efforts are concentrated on the replacement of C$_{2}$H$_{2}$F$_{4}$ using a gas known as \textit{tetrafluoropropene} (C$_{3}$H$_{2}$F$_{4}$) diluted with other gases to lower the detector working voltage \cite{prelGiorgia,prelAntonio,prelGianluca,prelPiccolo}. Promising gas mixtures have been identified by different research groups and their complete characterization in controlled data-taking environments (such as beam tests), as well as the study of the detectors long-term behavior when operated with eco-friendly gas mixtures is now needed.

To this aim, the RPC ECOgas@GIF++ collaboration (between ALICE, ATLAS, CMS, SHiP/LHCb and the CERN EP-DT group) was created to join forces among RPC experts of the different LHC experiments to share knowledge and manpower. Each group provided a detector prototype to be installed in a common support and the results shown hereafter have been obtained with the detector provided by the ALICE RPC group. 

This contribution is divided as follows: Section \ref{sec:experimental} contains a description of the experimental setup and the methodology used in the data-taking, Section \ref{sec:results} reports the main results obtained from a beam test campaign where mixtures with different HFO/CO$_{2}$ ratios have been beam-tested. Lastly, Section \ref{sec:conclusion} is dedicated to the conclusion and to possible outlooks for the future of this work.

\vspace{-20pt}

\section{Experimental setup}
\label{sec:experimental}

The experimental setup is located at the CERN Gamma Irradiation Facility (GIF++) \cite{GIF++}, installed on the H4 secondary beam line of the CERN SPS. This facility is equipped with a high activity ($\sim$12.5~TBq) $^{137}$Cs source and, in dedicated beam time periods, it is exposed to a high energy ($\sim$150~GeV) muon beam. The source can either be used to produce a high background radiation on the detector, simulating long operation periods in shorter time-spans (years in $\sim$months) or it can be combined with the muon beam to study the detector performance under varying background levels.

\begin{figure}[h!]
\includegraphics[width=\linewidth]{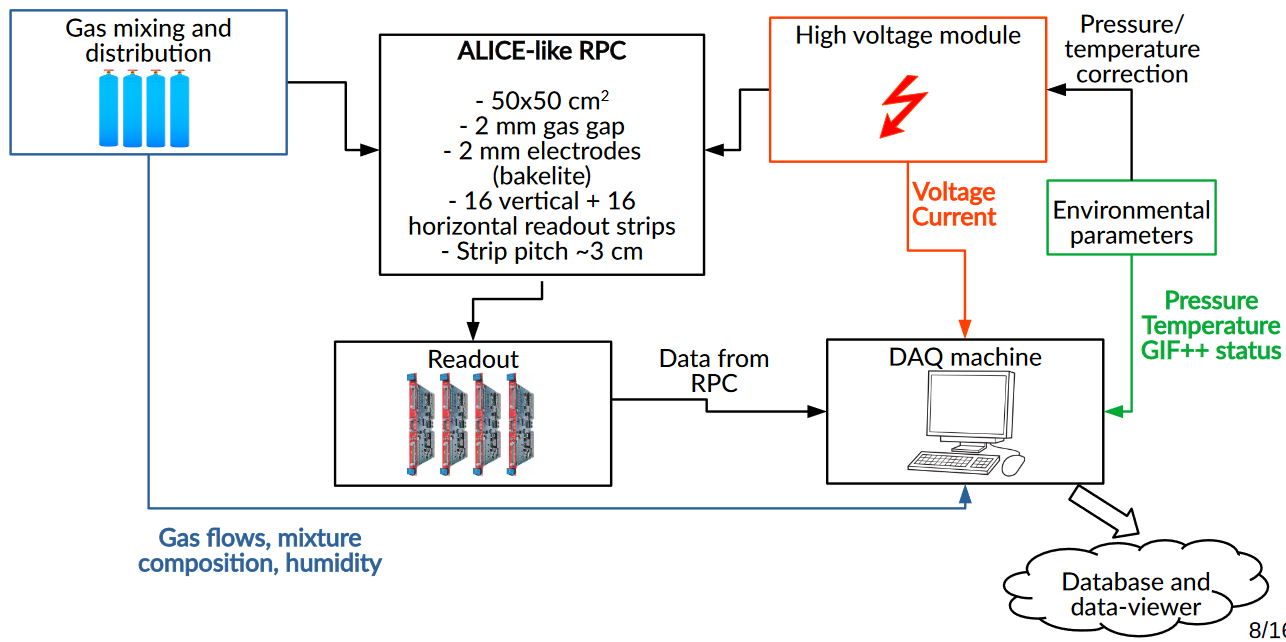}
\caption{Sketch of the experimental setup installed in GIF++ (a detailed description of each component is reported in the text)}
\label{fig:setup}
\end{figure} 

The $\gamma$ radiation can be modulated by means of lead filters and a total of 27 combinations can be obtained. The attenuation is measured with a pure number, which ranges from 1 to 46000 (maximum and minimum irradiation respectively). Moreover, the source can be fully shielded, allowing one to access the irradiation zone and to study the detector behavior without background radiation.

The ALICE detector is 50x50~cm$^{2}$ ALICE-like RPC prototype (made with the same construction process as well as with the same features as a real-life MID RPC). It is equipped with two perpendicular planes of 16 strips each (with a pitch of $\approx$3~cm). Figure \ref{fig:setup} reports a sketch of the experimental setup. Its main components are: the gas mixing and distribution system (4 Mass Flow Controllers to regulate the flow of each gas mixture component and a humidifier to add water vapour to the mixture, up to 40\% relative humidity), the high voltage system (a CAEN SY1527 mainframe\footnote{\url{https://www.caen.it/subfamilies/mainframes/}} and A1526P\footnote{\url{https://www.caen.it/products/a1526/}} high voltage board), the readout system, which can either be VME TDCs\footnote{V1190 with 128 input channels and time resolution of 100~ps} coupled with the ALICE MID front-end discriminators (FEERIC) \cite{FEERIC} or a CAEN digitizer\footnote{DT5742, 16 input channels, 12~bit resolution and 1, 2.5 and 5~Gs/s sampling rate} directly connected to the readout strips. Lastly, the DAQ machine supervising the whole data-taking procedure (as well as the continuous logging of relevant environmental parameters) is also shown in the figure.

A total of seven eco-friendly candidate mixtures has been tested, together with the mixture currently employed by CMS for reference. Table \ref{tab:mixtures} reports the composition of the different beam-tested gas mixtures (MIX0-6 for the eco-friendly and STD for the reference one).

\vspace{-15pt}

\begin{table}[h!]
	\begin{center}
		\setlength{\tabcolsep}{1.1pt} 
        \begin{tabular}{cccccc}\\\toprule  
        \textbf{Name} & \textbf{C$_{2}$H$_{2}$F$_{4}$ \%} & \textbf{HFO \%} & \textbf{CO$_{2}$ \%} & \textbf{i-C$_{4}$H$_{10}$ \%} & \textbf{SF$_{6}$ \%} \\\midrule
        STD & 95.2 & 0 &  0 & 4.5 & 0.3 \\  \midrule
        MIX0 & 0 & 0 & 95 & 4 & 1 \\  \midrule
        MIX1 & 0 & 10 & 85 & 4 & 1\\  \midrule
        MIX2 & 0 & 20 & 75 & 4 & 1\\  \midrule
        MIX3 & 0 & 25 & 69 & 5 & 1\\  \midrule
        MIX4 & 0 & 30 & 65 & 4 & 1\\  \midrule
        MIX5 & 0 & 35 & 60 & 4 & 1\\  \midrule
        MIX6 & 0 & 40 & 55 & 4 & 1 \\ \bottomrule
        \end{tabular}
        \caption{Composition of the gas mixtures used in the tests described}\label{tab:mixtures}
	\end{center}
\end{table}

\vspace{-35pt}

\section{Results}
\label{sec:results}

Section \ref{subsub:sourceOFF} describes the RPC response to the muon beam in the absence of $\gamma$ background (source-off) while Section \ref{subsub:sourceON} reports the results for varying levels of $\gamma$ background (source-on)

\subsection{Source-off results}
\label{subsub:sourceOFF}

\begin{figure} [h] 
    \centering
    \includegraphics[height=0.65\linewidth]{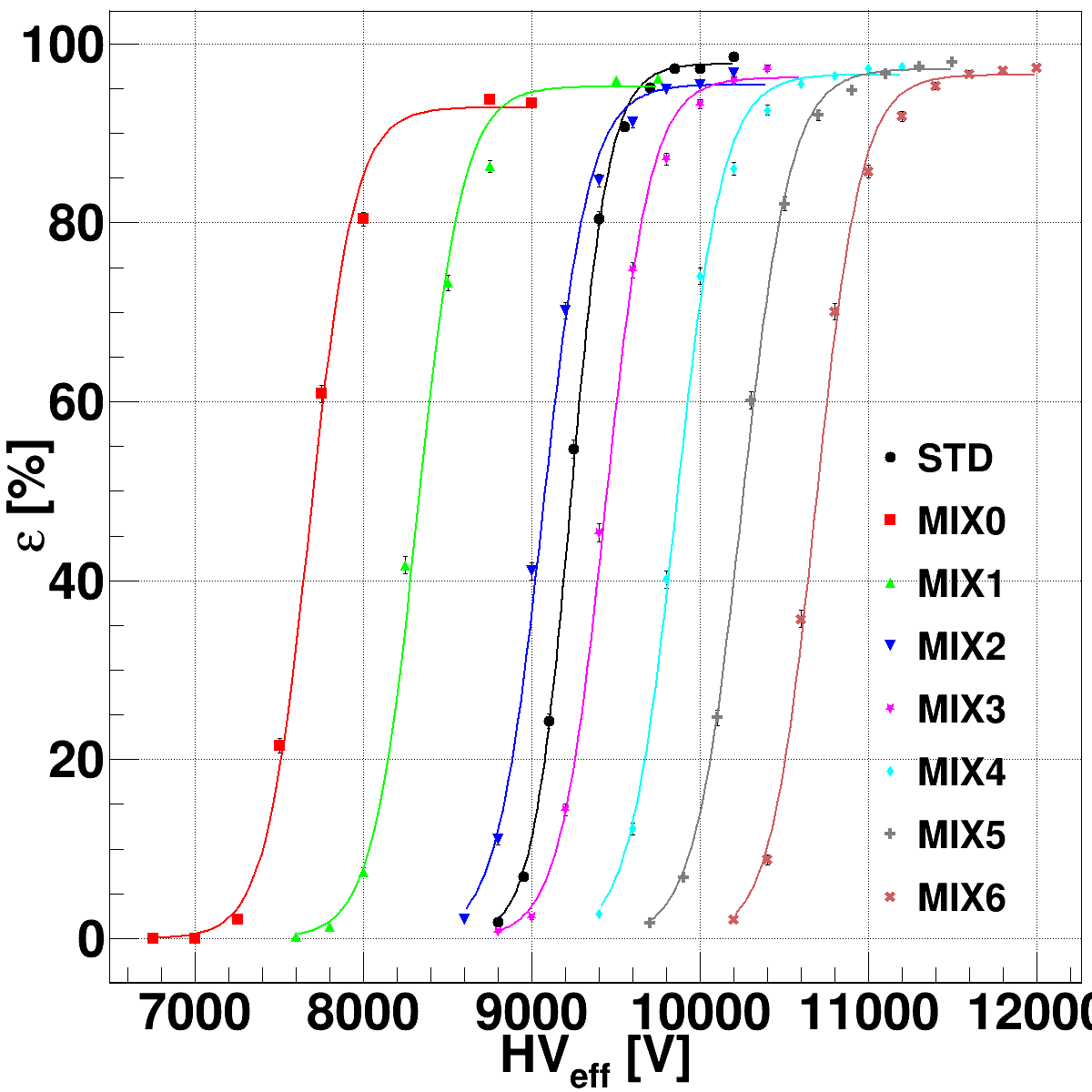}
    \caption{Efficiency as a function of the effective high voltage without gamma background. Results obtained when the readout strips are coupled directly to the digitizer}
    \label{fig:effDigit}
\end{figure}

\begin{figure} [h] 
    \centering
    \includegraphics[height=0.7\linewidth]{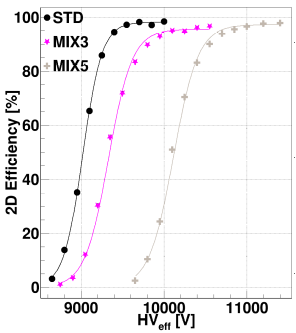}
    \caption{Efficiency as a function of the effective high voltage without gamma background. Results obtained when the readout strips are connected to the FEERIC discriminators}
    \label{fig:effFeeric}
\end{figure}

The first parameter that has been studied in the beam test is the RPC efficiency without any gamma background from the $^{137}$Cs source. This value has been computed using both the digitizer and the FEERIC front-end electronics. Figure \ref{fig:effDigit} shows the efficiency curves obtained when the digitizer was used. In this case, the threshold was set on the signal amplitude. A time interval where no muon signal is expected was identified as the "noise window" and the RMS of the signal amplitude was calculated in this interval. Then, if the signal amplitude, in a time region where a muon-induced signal is expected, is above five times the RMS calculated earlier(which usually translates to $\approx$1.5~mV), the RPC is considered efficient. Figure \ref{fig:effFeeric} reports the efficiency trend when the FEERIC front-end electronics are used in combination with the VME TDCs (in this case only the STD and two of the seven mentioned eco-friendly mixtures have been tested), the threshold was set to $\approx$150~fC before amplification. In general, the efficiency is calculated as the ratio between the number of signals above the thresholds and the total number of muon triggers.

\vspace{-20pt}

Note that the efficiency is shown as a function of the \textit{effective} high voltage (HV$_{eff}$), i.e. the voltage corrected for temperature and pressure variations, to keep the gain constant, according to the formula reported in \cite{CMSWP}.

In both cases, the efficiency curves have been interpolated using the logistic function, reported in Equation \ref{eq:logistc}

\begin{center}
    \begin{equation}
        \epsilon(HV) = \frac{\epsilon_{max}}{1+e^{-\lambda}(HV-HV_{50})}
        \label{eq:logistc}
    \end{equation}
\end{center}

where $\epsilon_{max}$ represents the maximum asymptotic efficiency reached by the detector, $\lambda$ is related to the steepness of the curve and HV$_{50}$ is the voltage where the efficiency reaches 50\% of its maximum. These parameters are used to extract the working point (WP, i.e. the voltage which grants full efficiency) as WP=$\frac{log(19)}{\lambda}$+HV$_{50}$ + 150~V, according to \cite{CMSWP}. 

By looking at both Figures \ref{fig:effDigit} and \ref{fig:effFeeric}, one sees that if the HFO concentration increases, so does the WP ($\approx$1~kV for every 10\% HFO added to the mixture). Moreover, the plateau efficiency reaches lower values if the HFO concentration is below 20\%. This could probably be explained by the lower number of primary electron-ion pairs produced by a MIP crossing CO$_{2}$ with respect to crossing C$_{2}$H$_{2}$F$_{4}$ \cite{Sauli}.

\begin{figure} [h] 
    \centering
    \includegraphics[height=0.7\linewidth]{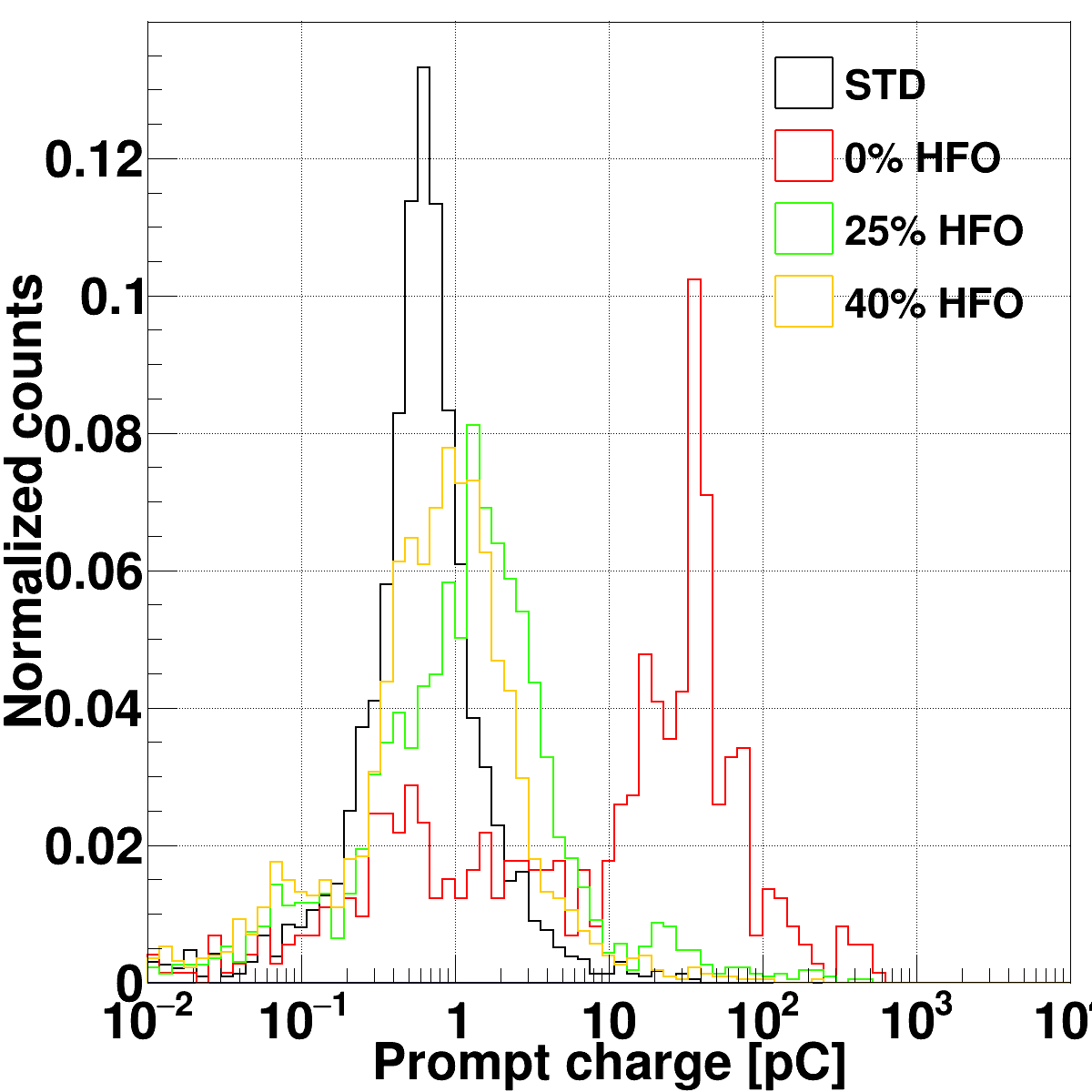}
    \caption{Prompt charge distributions at the HV$_{eff}$ closest to the estimated WP for selected gas mixtures}
    \label{fig:chargeWPsourceOFF}
\end{figure}

Another useful quantity, which can be computed when the RPC analog signals  are directly sent to a digitizer, and the full waveform of each hit is saved, is the signal prompt charge. This quantity is the time integral of the waveform and it has been computed following the procedure described in \cite{phdLuca}. Figure \ref{fig:chargeWPsourceOFF} shows the distribution of this quantity at the HV$_{eff}$ closest to the WP estimated earlier.

\vspace{-10pt}

In general, two distinct populations (separated at the $\approx$20~pC) can be observed; the one below 20~pC represents the avalanche signals (with smaller charge content) while the one above refers to the, so-called, streamers (larger signals due to avalanche degeneration). The STD gas mixture (in black) shows a single, well-defined peak and almost no streamers. For what concerns the eco-friendly mixtures, one sees that for lower HFO concentrations the avalanche peak is shifted towards higher values and the amount of streamers is higher (this translates to an increase of the absorbed current of $\approx$1.6/1.7 times with respect to the STD gas mixture). If the HFO concentration is increased (going from MIX0 to MIX6) the avalanche peak moves towards smaller values and the number of streamers also decreases, this is due to a quenching effect of more HFO in the mixture.

\subsection{Source-on results}
\label{subsub:sourceON}

The study of the RPC response to the muon beam, when exposed to a high intensity photon flux from the $^{137}$Cs source, is useful to simulate real-life operating conditions at the LHC. The expected background rate for the MID RPCs, during LHC Run 3 and 4, is $\approx$100~Hz/cm$^{2}$ \cite{Ferretti}, for the most exposed RPC during the Pb-Pb data-taking periods. 

\begin{wrapfigure}{r}{0.58\linewidth}
  \begin{center}
    \includegraphics[height=\linewidth]{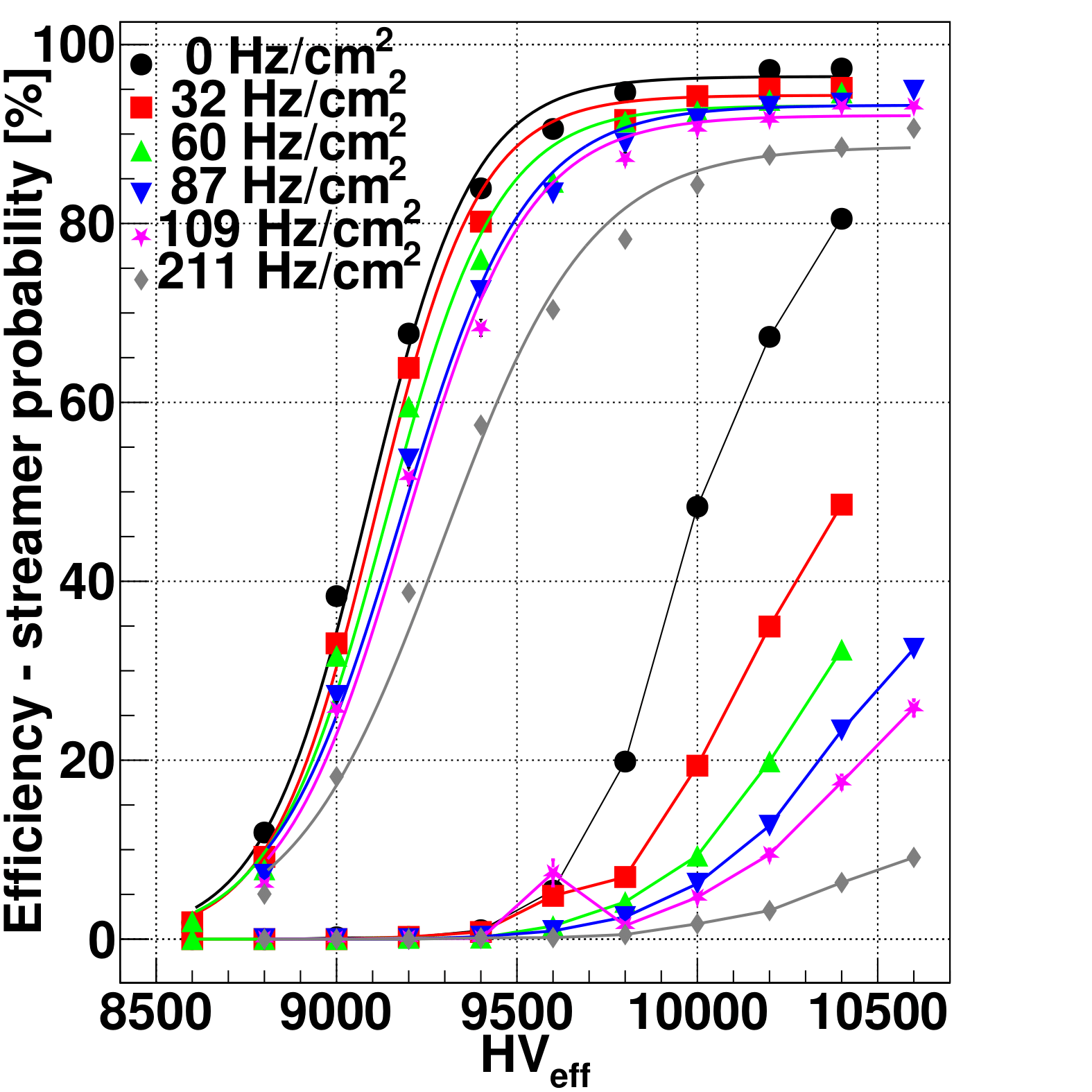}
  \end{center}
  \caption{Efficiency and large signal contamination as a function of the effective high voltage for increasing $\gamma$ background rates for the MIX2 mixture}
  \label{fig:effSourceOnNoCorr}
\end{wrapfigure}

Figure \ref{fig:effSourceOnNoCorr} shows the efficiency (measured using the digitizer), as well as the streamer contamination (calculated as the ratio between the number of signals with a charge above 20~pC and the total number of signals), as a function of HV$_{eff}$ for the MIX2 mixture, at increasing levels of $\gamma$ background.

Three main effects can be pointed out: if the $\gamma$ background increases, the efficiency curves tend to shift towards higher voltages, the plateau efficiency decreases and the streamer contamination is reduced. A possible explanation for this effect is that, when the $\gamma$ background increases, so does the current absorbed by the RPC. This current flows through all the detector elements, including the resistive electrodes. When it does so, it leads to an Ohmic voltage drop across them, leading to a decrease of the high voltage effectively applied to the gas, hence to a lower gain. All the three effects highlighted earlier seem to point to a gain decrease.

Since the current-induced voltage drop is Ohmic, one can calculate it as the product between the resistance of the electrodes and the flowing current. The latter is measured during data-taking while the former can be estimated using the Ar method \cite{Ar}. Using those quantities one can calculate the actual HV applied to the gas as: HV$_{gas}$ = HV - R\small$\times$\normalsize I where HV is the high voltage measured, R is the electrodes resistance and I the flowing current. 


\begin{figure} [h] 
    \centering
    \includegraphics[height=0.7\linewidth]{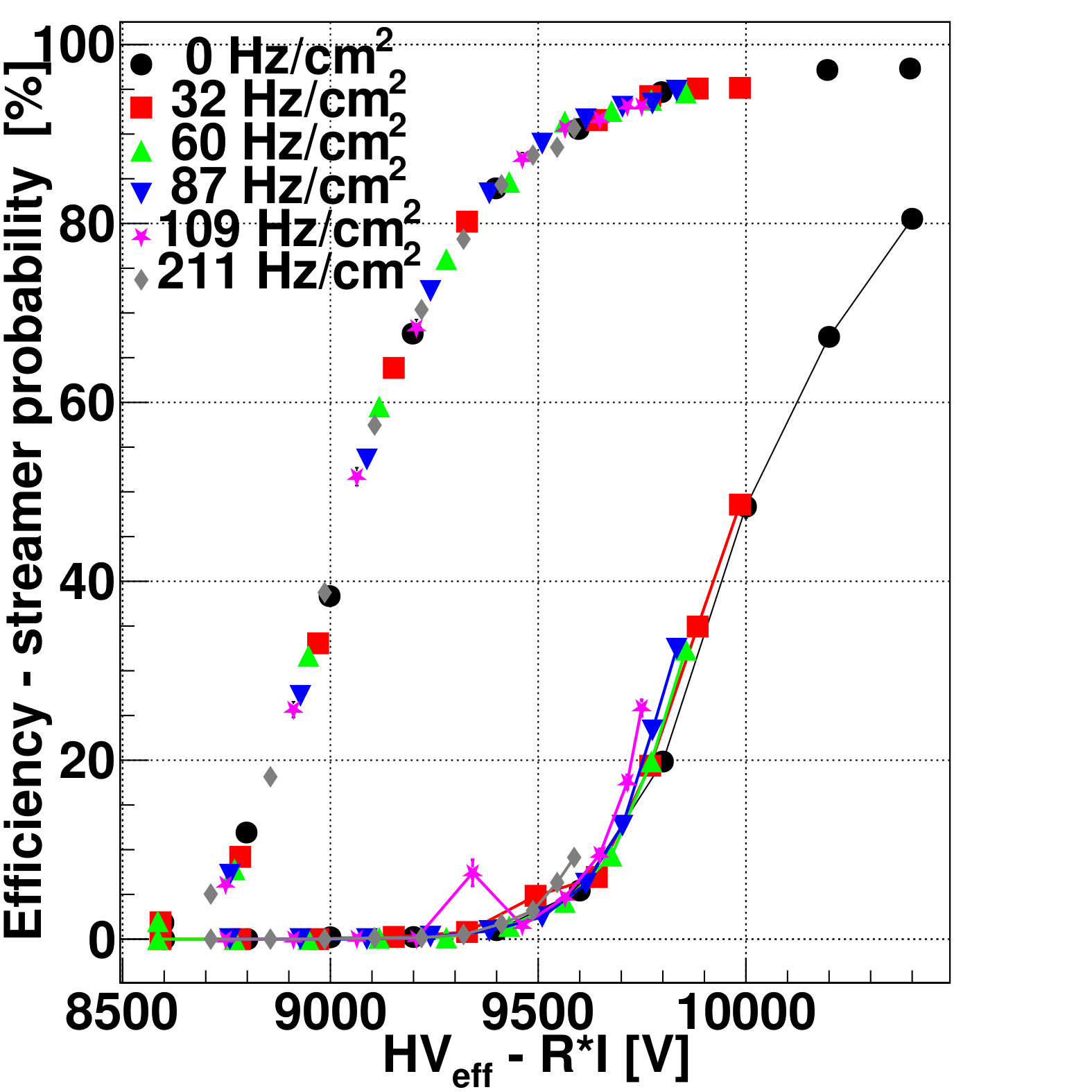}
    \caption{Efficiency and large signal contamination as a function of actual high voltage applied to the gas, for increasing $\gamma$ background rates for the MIX2 mixture}
    \label{fig:effSourceOnCorr}
\end{figure}

Figure \ref{fig:effSourceOnCorr} shows the efficiency and large signal (streamer) probability as a function of of HV$_{gas}$ and one can see how the curves are now superimposed (the same is true also for other mixtures, as reported in \cite{phdLuca}).


\section{Conclusions and outlook}
\label{sec:conclusion}

Several eco-friendly mixtures, where  C$_{2}$H$_{2}$F$_{4}$ is replaced with different ratios of HFO/CO$_{2}$ have been beam-tested at the CERN GIF++. First of all, the working point of the detectors increases by $\sim$1~kV for every 10\% HFO added to the mixture; secondly, the large signal contamination decreases if the HFO concentration increases; mixtures with 30-40\%~HFO provide values of streamer contamination similar to the ones given by the STD gas mixture, but streamers grow faster for voltages above the WP, reducing the safe operating region.

When exposed to gamma background, the efficiency curves shift to higher voltages and the maximum efficiency decreases (these observations can be explained by the current-induced voltage drop on the bakelite electrodes). In particular, for ALICE-like background rates ($\approx$100~Hz/cm$^{2}$) the efficiency drop is $\sim$1~percentage point (pp) for the standard gas mixture and $\sim$3-4 pp for the eco-friendly alternatives.

The results obtained from the beam tests are promising and a long-term aging test is now ongoing at GIF++, in order to study the stability of the detector response when operated for a long time with eco-friendly alternatives (preliminary results of these studies are reported in \cite{focusPoint}).





\end{document}